# Pressure-driven relaxation processes in nanocomposite ionic glass LiFe$_{0.75}$V$_{0.10}$PO$_4$

Szymon Starzonek[1*], Sylwester J. Rzoska[1], Aleksandra Drozd-Rzoska[1], Michał Boćkowski[1],

Tomasz K. Pietrzak[2] and Jerzy E. Garbarczyk[2]

[1]*Institute of High Pressure Physics of the Polish Academy of Sciences, Sokołowska 29/37, 01-142 Warsaw, Poland*

[2]*Faculty of Physics, Warsaw University of Technology*, Koszykowa 75, 00-662 *Warsaw, Poland*

*The Author to whom correspondence should be addressed: starzoneks@unipress.waw.pl

Abstract

This paper presents results for systems formed in a solid glassy state after nanocrystallization process above the glass temperature. We analyze electric conductivity and relaxation processes after such treatment under high temperature (HT) and high pressure (HP-HT) as well. The latter leads to ca. 8% increase of density, two decades ($10^2$) increase of electric conductivity as well as qualitative changes in relaxation processes. The previtreous-type changes of the relaxation time on cooling is analyzed by the use of critical-like and the 'critical-activated' description. Presented results correspond well with obtained for this material and shown in ref. [8]. The evidence for pressure evolution of the glass and crystallization temperatures, indicating the unique possibility of maxima and crossovers is also reported.

Introduction

The search for new generation materials for electric energy storage is driven by the boosted development of new functionality devices, possible only due to efficient and small-size batteries. The next grand motivation constitutes the global warming threat and the necessity of moving away from fossil fuels. The latter also means innovative batteries supporting the transformation of transport to electric vehicles and the construction of electricity storage facilities necessary for the complete introduction of renewable energy sources. A gradual crossover was expected in the forthcoming decades. However, the invasion of Russian troops on Ukraine in March 2022 caused the energy transformation to occur as fast as possible, becoming the biggest ongoing challenge.

Researchers studying energy storage suddenly find themselves on a peculiar front line of struggle with one of the greatest, even existential, threats to civilization. Formally, a battery is a relatively simple device composed of a cathode injecting ions, an anode and an electrolyte in between. The role of the cathode is crucial and mainly responsible for efficiency, working durability, and costs. In the last decade, glassy $LiFePO_4$ based materials have been indicated as the base for durables, a cost-effective and hypothetically efficient reference for cathode materials. Glasses have an 'open chaotic structure' with a large free volume. They are also characterized by decoupling between ionic, electronic conduction, and structural mode relaxations by decoupling between ionic conduction and structural mode relaxation [1-6].

Notwithstanding, despite a decade of studies, relatively low electric conductivity remains the main drawback against its ultimate application. Several approaches have been implemented to overcome this fundamental problem. One of the paths considered is the vanadium-related supplementation of the molecular structure, for which a positive influence on electric conductivity and electrochemical properties is expected [4, 5]. Another way to improve desired properties is by creating micro/nano- crystalline centers to facilitate electrical conductivity via the Mott hoping mechanism. Some of the authors of this report developed the approach based on creating olivine-type nano-centers within the amorphous glassy matrix via the short-time thermal nanocrystallization just above glass transition temperatures. In particular, Garbarczyk et al. [4, 5] developed the scenario associated with rapid heating followed by a subsequent quench which led to the formation of the composite nano-crystallites system to LiFeV$_{0.10}$PO$_4$. They noted strong increase of electric conductivity. The phenomenon was explained by adopting the classical Mott model., namely Mott's theory of polaron hopping between nanocrystalline centers. The model yields the following output relation for DC electric conductivity $\sigma$ [4, 5, 7]:

$$\sigma(T) = \nu_e(1-c)\frac{e^2}{Rk_BT}\exp(-2\alpha R)\exp\left(-\frac{E_a}{k_BT}\right) \quad (1a)$$

$$E_a = \frac{e^2}{16\pi\varepsilon_0\varepsilon_P r_p}\left(1+\frac{r_p}{R}\right)+\frac{W_D}{2} \quad (1b)$$

where $E_a$ denotes the activation energy, $R$ is the average distance between hoping centers, $\alpha$ is the inverse of the localization length electron wave function, $c$ is the fraction of hoping centers in the system, $\varepsilon_P$ is the effective electric permittivity of polarons and $r_P$ is for its radius; $W_D$ stands for the average difference between nearest hoping centers potentials.

As a result of the above relation, the electric conductivity, including the governing activation energy, strongly depends on the distance $R$ and the 'concentration' $c$. Compressing can change both properties, which shows the notable impact already in the 'low GPa' domain for glassy materials. It offers an interesting path to support the achievement of favorable electrical conductivity due to the subjecting of the discussed homo-nanocomposite material to high pressures [8]. Unfortunately, there is a fundamental problem here observed in many materials subjected to high pressure: its influence can qualitatively change properties, inducing even exotic states with unusual properties. After decompressing, there is a return to the initial state, so the pressure experiment becomes more of an indication of the discovery of unique properties. One can only hope that 'some-day, in 'some-way' will be regained under normal conditions [9, 10]. However, in the case of the transition to the glass state, it was found that high-temperature and high-pressure annealing in the magic domain resided in the solid glass phase between ca. $0.9T_g$ a $T_g$ can induce special properties that are permanently retained after decompression and recovery [10-16]. It is worth emphasizing that such changes are obtained for $P \sim 1$ GPa, and even below. It can be explained by the link vitreous materials with the category of Soft Matter, which various types showed extreme sensitive to external disturbances, preserved after decompressing [17-20]. The mentioned extraordinary possibilities of forming properties with the use of HP-HT annealing were discovered in the study of silicate and oxide glasses, where record surface hardness or densification approaching even 20% of the initial value was achieved [10-16]. The last feature obviously must lead to a significant increase of the refractive index ($n$) or dielectric constant ($\varepsilon$), due to the link between density ($\rho$), $n$, expressed in the Mossotti-Clausius-Lorentz local field [21], which seems to describe the properties of the solid glasses phase quite well [22].

Following the path proposed by the authors in ref. [8], the research results were presented, which was LiFe$_{0.75}$V$_{0.10}$PO$_4$, which was subjected to HP-HT transformation using the above-mentioned remarks, obtaining a significant and permanent increase in electric conductivity, which was shown for measurements in the range from 373 K to 473 K. The work mentioned above also includes basic material characteristics (XRD, DSC) and shows the linear pressure changes of the glass temperature $T_g(P)$ and the above-average temperature characterizing the local crystallization zone $T_c(P)$. In the liquid phase supercooled above $T_g(P)$: these tests were performed down to 1 GPa.

The results presented in this paper qualitatively expand the images of the LiFe$_{0.75}$V$_{0.10}$PO$_4$ homo-nanocomposites formed after HP-HT annealing, mainly showing fundamental changes in the map of relaxation processes, an extremely important and rarely discussed property. The scope of the DC conductivity comparative analysis was extended up to the 400 K wide region, which revealed its non-linear characteristic on the Arrhenius scale. Finally, the range of the pressure characteristics of $T_g(P)$ and $T_c(P)$ was extended to record values above 2 GPa. The latter research was motivated by the authors 'discovery of non-linear $T_g(P)$ changes in pure' LiFePO$_4$, including the remarkable crossover of $dT_g/dP > 0 \Rightarrow dT_g/dP < 0$ on compressing, occurring below 1 GPa [23].

## Experimental

Experimental details regarding the preparation and treatment of LiFe$_{0.75}$V$_{0.10}$PO$_4$ are described in detail in ref. [8], which also includes XRD and DSC characterization of samples. DSC scans were carried out under pressure, in ref. [8] reaching 1 GPa and in the given report 2.2 GPa. Notable is the number of tested samples reaching $m \sim 5g$, which results in relatively large temperature changes associated with detected processes. The broadband dielectric spectroscopy (BDS) studies were carried out using the Novocontrol (1µHz-10MHz) impedance analyzer. The results presented are focused on DC electric conductivity for the low frequencies focus. The enhanced range of frequencies employs the complex modulus representations analysis, appropriate for mapping relaxation processes in conductive samples.

## Results and Discussion

For the formation of nanocomposite ionic glass LiFe$_{0.75}$V$_{0.10}$PO$_4$ using the HP-HT treatment, necessary is the knowledge regarding pressure dependences of the glass temperature $T_g$ and crystallization $T_{cr.}$ zone. In ref. [8] it was determined up to $P = 1$ GPa, showing linear changes for both properties. It was obtained via DTA studies in situ under pressure, using large volume (*m* ~ 5g) samples. Figure 1 presents the results of the extension of these studies up to pressures well above 2 GPa, with the nonlinearity emerging on compressing. The pressure evolution of the melting/crystallization temperature $T_{cr}$ is most often parameterized via the Simon-Glatzel equation (1929, [24], SG), and by its ad hoc introduced parallel by Andersson & Andersson (1998, [25], AA) for the glass temperature [26]:

$$T_{g,m,cr.}(P) \approx T_{g,m}^0 \left(1 + \frac{P}{a}\right)^{1/b} \qquad (2)$$

where *a* and *b* denote empirical coefficients and the reference temperature $T_g^0 = T_g(P = 0) \approx T_g(P = 0.1 MPa)$.

SG and AA relations predict the permanent $T_{g,m}(P)$ rise on compressing, i.e., $dT_{g,m,cr.}/dP > 0$ However, for melting, there is a notable group of systems where the maximum of $T_m(P)$ occurs [26-28]. The experimental evidence for $T_g(P)$ is still very limited, what experimental problems can explain [26]. Notwithstanding, worth recalling is the Turnbull criterion linking both temperatures: $T_m/T_g \approx 2/3$ [29], which may suggest their parallel changes of $T_m(P)$ and $T_g(P)$. Eq. (2) can be easily validated for the given set of experimental data, using the following transformation [30-32]:

$$\left(\frac{d\ln T_g(P)}{dP}\right)^{-1} = ab + bP = A + BP \qquad (3)$$

where $A, B = const$

The description of $T_{g,m}(P)$ by Eq. (2) is valid if a linear dependence appears for the check-in plot defined by Eq. (3). In refs. [30-32] the relation portraying melting and vitrification in systems with the crossover $dT_{g,m}/dP > 0 \Rightarrow dT_{g,m}/dP < 0$, namely:

$$T_g(P) = F(P)D(P) = T_g^0 \left(1 + \frac{\Delta P}{\Pi}\right)^{1/b} \exp\left(-\frac{\Delta P}{c}\right) \qquad (4)$$

where $F(P)$ and $D(P)$ are for the rising (SG-type) and damping terms; $-\pi < 0$ is for the terminal absolute stability limit pressure at $T = 0$, $\Pi = \pi + P_g^0$, $\Delta P = P - P_g^0$, and $c$ is for the damping pressure coefficient.

For this relation, the following check-in analysis of experimental data was proposed [30-32]:

$$\left(\frac{dlnT_g(P)}{dP} + c^{-1}\right)^{-1} = A + BP = b\pi + b\Delta P \tag{5}$$

The linear regression fit based on linear evolution defined by Eq. (5) yields optimal values of parameters for Eq. (4). It is pressure-invariant, i.e., it can be applied for the arbitrary reference temperature and pressure $\left(T_{g,m}^0, P_{g,m}^0\right)$ along $T_{g,m}(P)$ curves. The analysis via Eqs. (4) and (5) can yield the same values of parameters for any tested pressure domain within $T_{g,m}(P)$ experimental data, enabling the reliable extrapolation.

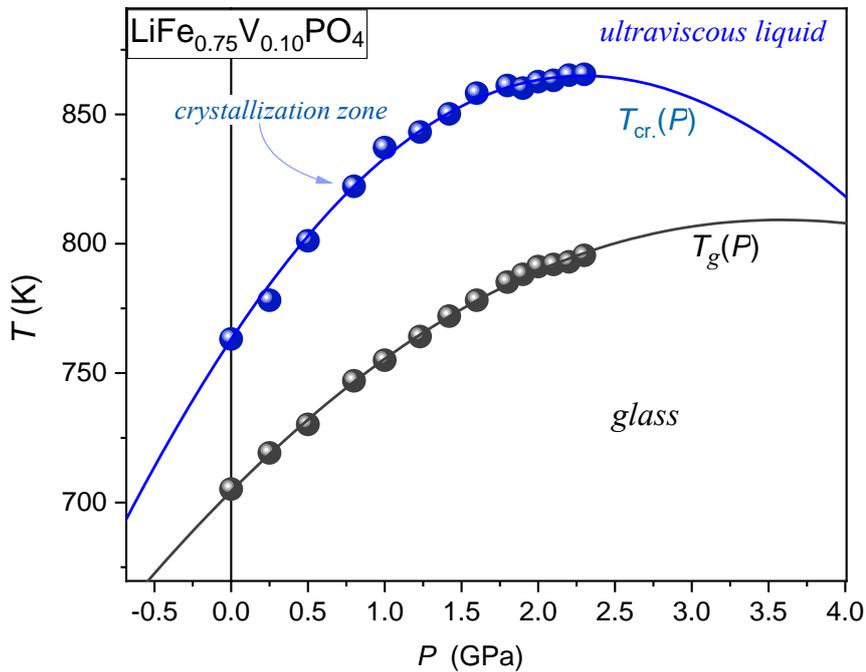

FIGURE 1 Pressure depends on the glass temperature and the crystallization zone temperature in the supercooled liquid state for LiFe$_{0.75}$V$_{0.10}$PO$_4$. Both evolutions are portrayed by Eq. (2).

Solid curves in Fig. 1 portraying experimental data are related to Eq. (3), with parameters obtained via the supporting derivative analysis using Eq. (5). The latter is shown in the inset in Figure 2. The central part of this plot is related to Eq. (3), and illustrates the inadequacy of AA and SG (Eq. 2) for portraying experimental data in Fig. 1. The notable feature resulting from the above description of experimental data is the decrease in distance between $T_g$ and $T_m$ both on high compressing and entering the negative pressures domain, related to isotropic stretching

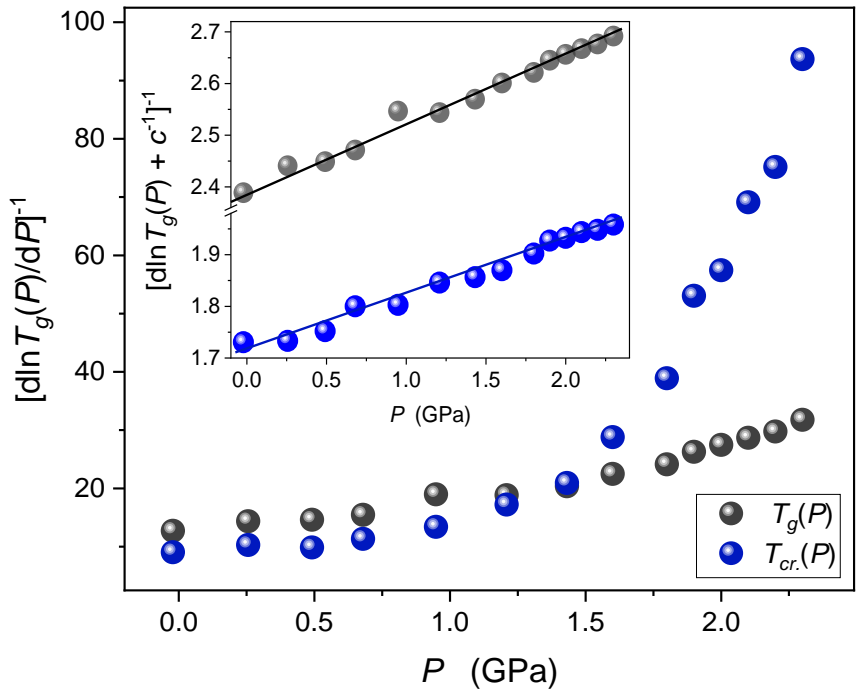

FIGURE 2 Results of the derivative-based analysis of T_(g,m) (P) experimental data from Fig.1, via Eq. (3) and Eq. (4), focused on the validation (manifesting via linear changes) of SG-AA Eq. (2) and ADR Eq. (4), respectively.

TABLE I

Results of the analysis of $T_g(P)$, $T_{cr.}(P)$ via Eqs. (4 and 5).

| Parameters<br>Curve | $b$ | $-\pi$ (GPa) | $c$ |
|---|---|---|---|
| $T_{cryst.}(P)$ | 0.16 | 9.9 | 1.98 |
| $T_g(P)$ | 0.20 | 12.7 | 3.33 |

As mentioned above, the annealing of compressed glass samples in the solid state, just below the glass temperature, can also cause the preservation of the acquired specific properties after decompression. For the tested system, the essential meaning has the acquired DC electric conductivity. Such behavior in the extended temperature range exceeding 300 K is shown in Figure 3, for samples after nanocomposite formations after the classic HT and the 'innovative' HP-HT treatments. The inset is for the distortions-sensitive apparent activation enthalpy $H_a(T)$ insight. This magnitude is associated with the concept of the general Super-Arrhenius (SA) dynamics [32, 33]:

$$\tau(T), \sigma(T) \propto \exp\left(\frac{E_a(T)}{RT}\right) = \exp\left(\frac{E'_a(T)}{T}\right) \qquad (6)$$

where $E_a(T)$ is for the apparent (temperature-dependent) activation energy. For $E_a(T) = E_a = const$ in the given temperature domain, one obtains the basic Arrhenius portrayal. Taking the derivative of the above relation, one obtains [34-36]:

$$\frac{d\ln\tau(T)}{d(1/T)} = H_a = E'_a + \frac{1}{T}\frac{dE'_a}{d(1/T)} \qquad (7)$$

It is visible that the apparent activation enthalpy and the apparent activation energy coincide only for the basic Arrhenius case.

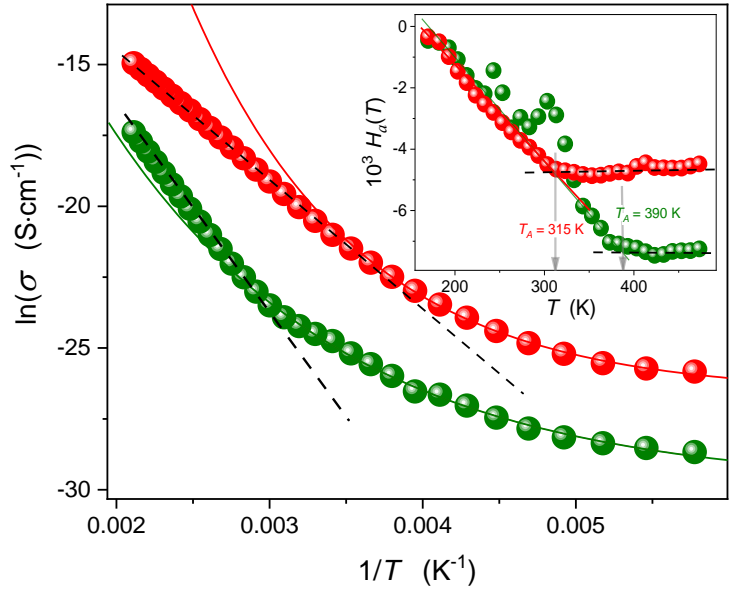

FIGURE 3 Temperature evolution of DC electric conductivity in solid after nanocrystallization process carried out under atmospheric pressure (green) and at P = 1 GPa (red), in LiFe$_{0.75}$V$_{0.10}$PO$_4$. The insight presents the distortions-sensitive, apparent enthalpy based analysis: $H_a(T) = dln\sigma(T)/d(1/T)$. Solid curves/lines are related to the non-Arrhenius behaviour and Eq. (8). Dashed lines are for the basic Arrhenius behaviour.

*Table II*
*Results of portraying experimental for the non-Arrhenius domain of electric conductivity changes in Figure 3, via Eq. 8. Graphically results are shown by solid curve sin this Figure.*

| Parameters<br>Process | $ln\sigma_0$ | $C'$ | $E$ |
|---|---|---|---|
| HT | -29.9 | 45.4 | -648.1 |
| HP-HT | -26.6 | 135.8 | -920.5 |

In Fig. 3 the Arrhenius behavior appears as the applied linear behavior for the scale. It obeys only the high-temperature limit. On cooling, the non-Arrhenius behavior dominates, but the decrease of the electric conductivity is much lesser than observed in glass forming systems. The inset in Figure 3 reveals the range of the basic Arrhenius behavior explicitly and shows the linear rise of $H_a(T)$ on cooling. They are smooth only after HP-HT treatment. Such behavior of $H_a(T)$ and the definition of the apparent activation enthalpy leads to the following portrayal of electric conductivity changes in the non-Arrhenius domain:

$$\ln\sigma(T) = \ln C + F\exp\left(-\frac{E}{T}\right) \Rightarrow \sigma(T) = C\exp\left[F\exp\left(\frac{E}{T}\right)\right] \approx C + C'\exp\left(\frac{E}{T}\right) \qquad (8)$$

Results of such portrayal are shown in Fig. 3, with parameters given in Table II.

Notable that the focus of the analysis on the reference Mott's model Eq. (1), i.e., the analysis using $T\ln\sigma(T)$ or $T\log_{10}\sigma(T)$ vs. $1/T$ scale, does not influence results.

In highly conductive systems, the detailed insight into relaxation processes, enabling even their decoupling, offers the complex dielectric modulus analysis [32, 33]. Examples of such spectra are presented in Figure 4. The temperature evolution of these relaxation times is presented in Figure 5. They reveal clear dielectric loss curves, in which peaks (maxima) define related relaxation times $\tau_{oi}$, $i = 1,2,3$. The latter is related to the emergence of 3 relaxation processes, at least in the HT sample. Only two processes remain after the high compression treatment (HP-HT sample). When comparing relaxation maps after HT and HP-HT treatments, striking is the reduction of relaxation processes for the latter, matched with the emergence of the Arrhenius - non-Arrhenius crossover for the latter, showing the 'previtreous slowing – down behavior'.

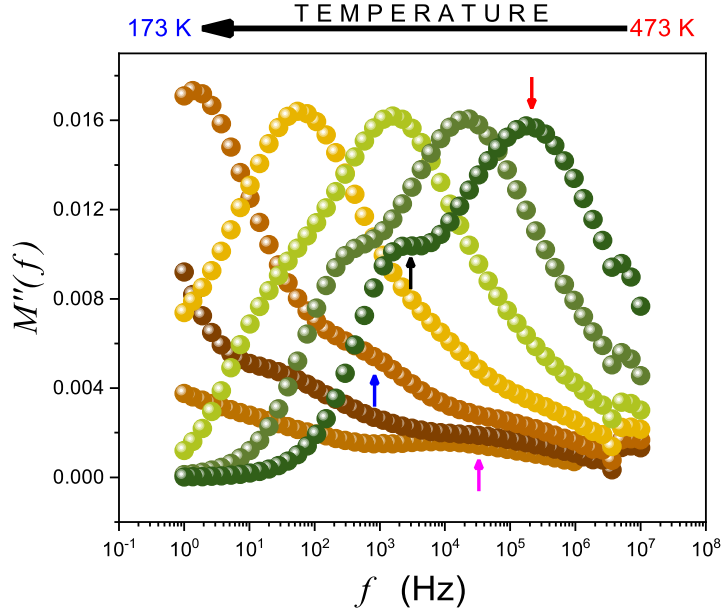

FIGURE 4 Temperature evolution of an imaginary part $M''(\omega)$ of the complex dielectric modulus $M^*(\omega) = M'(\omega) + iM''(\omega)$, $\omega = 2\pi f$ in LiFe$_{0.75}$V$_{0.10}$PO$_4$. Arrows show different relaxation processes as follows: ionic (↓), low-frequency (↑), beta (↑),

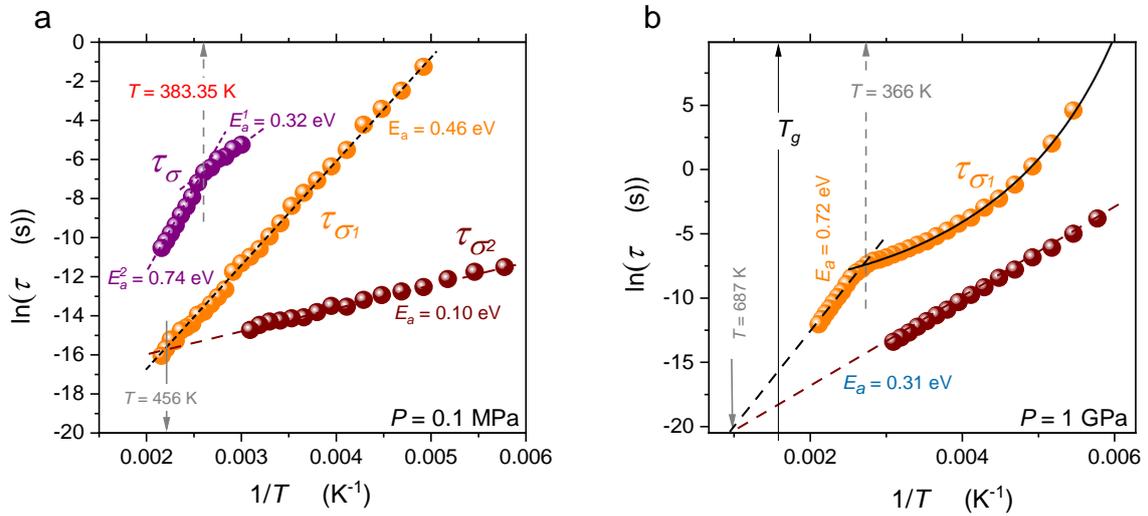

FIGURE 5 Relaxation maps for homo homo-nanocomposite after thermal ($P = 0.1 MPa$) and $P = 1 GPa$ formation of LiFe$_{0.75}$V$_{0.10}$PO$_4$ glassy nanocomposite. The latter pressure is associated with HP-HT annealing at $T = 0.95 T_g(P = 1GPa)$. The red curve in Fig. 1b is related to critical-activated Eq. 8, with the singular temperature $T^* = 137.5K$ e, exponent $\Omega = 21.6$ and prefactor $lnC_\Omega = -31.02$.

Figure 6 presents the apparent activation energy analysis of $\tau_{\sigma 1}(T)$. It reveals the superior via the critical-like dependence [37]:

$$H_a(T) = \frac{d\ln\tau(T)}{d(1/T)} \approx \frac{M'}{T-T^*} \tag{9}$$

The apparent activation enthalpy is directly related to the so-called apparent fragility $m_P(T)$, introduced in supercooled glass-forming systems to describe relative changes of the relaxation time on cooling towards the glass temperature $T_g$:

$$m_P(T) = \frac{d\log_{10}\tau(T)}{d(T_g/T)} = \frac{1}{T_g \ln 10}\frac{d\ln\tau(T)}{d(1/T)} = \frac{1}{T_g \ln 10}\frac{H_a(T)}{R} = \frac{1}{T_g \ln 10}H_a(T) \tag{10}$$

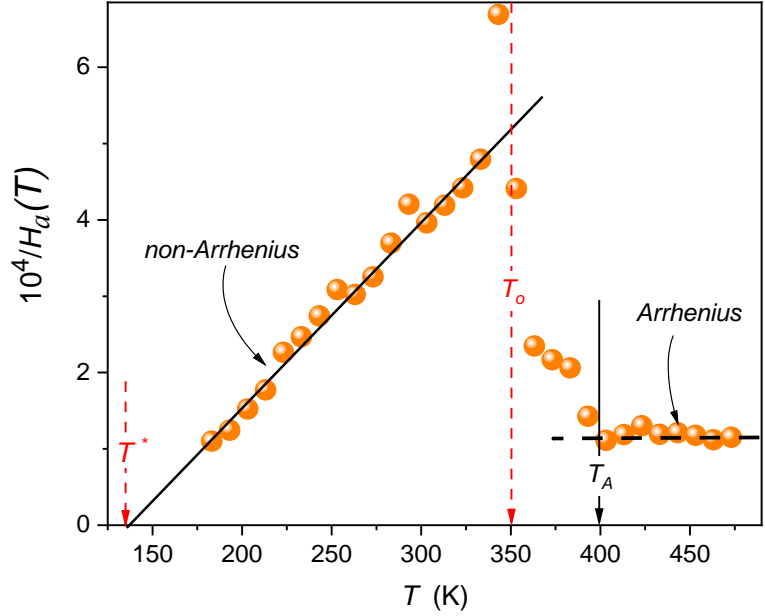

*FIGURE 6 The temperature behavior of the apparent activation enthalpy $H_a(T) = d\ln\tau_{\sigma 1}/d(1/T)$ in solid glass LiFe$_{0.75}$V$_{0.10}$PO$_4$, based on data presented in Fig. 5b . The horizontal dashed line is for the Arrhenius evolution, and the solid line is for the 'universal' behavior of $H_a(T)$ given by Eq. (8) and coupled to the 'critical-activated' Eq. (9).*

Notable, that the parallel of Eq. (9) also obeys for the pressure path of approaching the glass transition [38, 39]. In ref. [37] the empirical finding of the previtreous evolution of the apparent fragility parallel to Eq. (9) led to the derivation of the novel 3-parameters relation describing the previtreous dynamics, linking critical-like and activated (i.e., related to SA equation) behavior:

$$\tau(T) = C_\Omega \left[ \left(\frac{T-T^*}{T}\right)^{-1} exp\left(\frac{T-T^*}{T}\right) \right]^\Omega = C_\Omega [t^{-1} exp(t)]^\Omega \qquad (11)$$

where $t = (T - T^+)/T$ ; $C_\Omega$ and $\Omega$ are constants.

Results of experimental data fitting via Eq. (11) are shown by the solid, red curve in Fig. 5, with parameters given in the caption.

## Conclusions

Nanocomposite systems formed in a solid glassy system after 'limited' nanocrystallization above the glass temperature followed by fast cooling to ambient conditions is indicated as an effective treatment for increasing the electric conductivity, necessary for possible application as the cathode in new-generation batteries. LiFe$_{0.75}$V$_{0.10}$PO$_4$ constitutes here the key model exemplification of such system. This report analyzes electric conductivity and relaxation processes after such HT treatment with the formation under high temperature and high pressure (HP-HT). The latter leads to ca. 8% increase in density, two decades increase in electric conductivity, and qualitative modifications of the relaxation processes map, extending over 300 K. The HP-HT treatment is related to the 'fast' nanocrystallization above $T_g$, followed by 30 – 60 minutes of annealing in the solid glass state just below $T_g$ . The latter stage causes the preservation of specific features after decompressing and returning to ambient conditions. It does not occur if the annealing just below $T_g$ is omitted or carried out well below $T_g$, for instance at $T_g - 100K$. When comparing LiFe$_{0.75}$V$_{0.10}$PO$_4$ samples after HT ( $P =$ $0.1 MPa$) and HP-HT ($P = 1 GPa$) treatment, a significant increase in electric conductivity, essential for applications, is visible. The reduction of relaxation processes and the appearance of non-Arrhenius dynamics after high-pressure operations. Regarding the latter, the previtreous-type increase of the relaxation time on cooling is evidenced, for which the superior critical-like changes of the apparent activation energy and consequently the 'critical-activated' portrayal of relaxation time is evidenced. Similar behavior was obtained for the DC electric conductivity. These issues are worth stressing since, commonly, the portrayal of the non-Arrhenius dynamics is carried out via the Vogel-Fulcher-Tamman (VFT) replacement super-Arrhenius equation with the apparent activation energy: $\tau(T), \sigma(T) \propto exp(E_a(T)/RT)$, the exemplification of the 'activated' dynamics.

For the evolution of dynamic properties in the solid, glass stated the basic Arrhenius evolution pattern $(E_a(T) = E_a = const)$ is generally advised. This report shows the emergence of the non-Arrhenius previtreous dynamics in the solid glass after the HP-HT treatment, i.e., the unique creation of a supercooled-liquid-like behavior in solid glass 'matrix'. One can call it as '*glass in glass*' formation, probably associated with the specific formation of the hoping mechanism between nanocrystallites after HP-HT treatment. All of these can be important for applications recalled above but constitute a new phenomenon for the glass transition physics, not reported so far.

Finally, we would like to stress the evidence for pressure evolution of the glass and crystallization temperatures, indicating the unique possibility of maxima and crossovers $dT_{g,cryst.}/dP > 0 \quad \Rightarrow \quad dT_{g,cryst.}/dP < 0$ on compressing.